\documentclass[prl,twocolumn,floatfix]{revtex4}
\usepackage{graphics}
\usepackage{graphicx}
\usepackage{amsmath}
\usepackage{amstext}
\usepackage{xcolor}
\usepackage{}
\usepackage{amssymb}
\usepackage{hyperref}

\newcommand{\ex}[1]{\mathrm{e}^{#1}}

\newcommand{\im}{\mathrm{i}}

\newcommand{\vect}[1]{\boldsymbol{#1}}

\begin{document}
\title{Unusual states of vortex matter in mixtures of Bose--Einstein Condensates on rotating optical lattices}
\author{E. K. Dahl$^{1}$, E. Babaev$^{2,3}$ and A. Sudb{\o}$^{1}$}
\affiliation{
$^{1}$Department of Physics, Norwegian University of Science
and Technology, N-7491 Trondheim, Norway\\
$^{2}$Physics Department, University of Massachusetts, Amherst MA 01003, USA\\
$^{3}$Department of Theoretical Physics, The Royal Institute of Technology 10691 Stockholm, Sweden}

\date{\today }

\begin{abstract}
In a single-component superfluid under rotation  a broken 
symmetry in the order parameter space results in a broken translational symmetry in real space: a vortex lattice. 
If  translational symmetry is restored, the phase of the order parameter  disorders and 
thus the broken symmetry in the order parameter space is also restored. We 
show that for Bose-Condensate mixtures in optical lattices with negative dissipationless  drag, a new situation  arises. This 
state is a modulated vortex liquid which breaks translational symmetry in the 
direction transverse to the rotation vector.

\end{abstract}
\maketitle
{
An important property of a superfluid is its specific rotational 
response. Namely it comes into rotation by means of 
the formation of a vortex lattice.} Under the influence of other factors such as temperature, 
multiplicity of superfluid components, inhomogeneities etc., different ``aggregate'' states 
of vortex matter may form, such as vortex liquids, glasses, etc \cite{Glass}. The variety of states is 
even richer in multicomponent systems \cite{multicomp1}. 
The transitions between the various ``aggregate" states of vortex matter are related to
various ordering processess of particles in condensed matter systems. For example, 
the process of thermal vortex lattice melting can be mapped onto an insulator-to-superfluid 
transition of bosons. In this mapping, a vortex line is viewed as a world line of a boson with 
the $z$-axis mapped onto a ``time''-axis  and the vortex liquid, which is also entangled, 
represents the delocalized/superfluid state of the dual bosons \cite{MPAFisher}. The central 
result of the present work is that we find evidence in large-scale Monte Carlo (MC) computations 
that in a two-component Bose--Einstein condensate (BEC), the  vortex 
lines support a state possessing properties of a vortex liquid simultaneously with properties 
of vortex lattice i.e. breakdown of translational symmetry.
\par
Recent progress in creating and observing various {mixtures} of multi-component 
BEC has produced much interest in these systems. 
{ When BEC components 
are not spatially separated, the generic type of interaction between them is the 
current-current interaction  (Andreev--Bashkin effect)\cite{AB} describing non-dissipative 
drag between the two superfluid components.} { Such a system in units where $\hbar=1$ 
is described by the free energy density \cite{AB,Kuklov1,Dahl1}}
\begin{align}
\label{model}
F=\frac{1}{2}\bigg\{&m_1n_1\left(\frac{\nabla\theta_1}{m_1}-\vect{\Theta}\right)^2
+m_2n_2\left(\frac{\nabla\theta_2}{m_2}-\vect{\Theta}\right)^2\nonumber \\
&-\sqrt{m_1m_2}n_d\left(\frac{\nabla\theta_1}{m_1}-\frac{\nabla\theta_2}{m_2}\right)^2\bigg\},
\end{align}
{ where $m_1,m_2$, $\theta_1$ and $\theta_2$  are the masses and  the phases of the
condensates while $n_1,n_2$  control the phase stiffnesses of the two components.
Further, the drag coeficient $n_d$ controls the  density of one component  dragged by the superfluid velocity of the other 
component}.   Since here we are interested in the physics of rotating system we 
include the field $\vect{\Theta}$  which accounts for  rotation with angular velocity
$\vect{\Omega}=\nabla\times\vect{\Theta}=2\pi f\hat{\vect{z}}$, where
$m_if$ is the number of rotation-induced vortices per unit area of component $i$. 
{ We will use $f=1/64$ throughout}. In the 
following, we denote vortices with $2\pi l_{(i,j)}$ windings in $\theta_{(1,2)}$ with a pair of 
integers $(\Delta\theta_1=2\pi l_1,\Delta\theta_2=2\pi l_2)=(l_1,l_2)$. The last term in \eqref{model} 
is the current-current interaction\cite{AB} which may be caused by different reasons, such as 
intercomponent van der Waals interaction\cite{AB} or originating with the underlying optical lattice 
\cite{Kuklov1}. It was first considered in the contexts  of the physics of
${}^3$He -${}^4$He mixtures and coexisting neutronic and protonic condensates in neutron stars. 
Various aspects of the rotational response of this system has been studied so far only for 
{\it positive} values of drag in context of ${}^3He-{}^4He$ mixtures \cite{AB} and  BEC 
mixtures \cite{Dahl2}. 
{ However,
it has been recently shown that in  
optical lattices  there arises an 
intriguing possibility to produce a BEC mixture with  a {\it negative} inter-species drag 
$n_d$ \cite{Kuklov1}. }
\par
{ In this paper we address the
physics of a rotating system with a negative intercomponent drag
and find it  being very rich.}
This is manifested in the situations where the 
usual notions from disordered versus ordered vortex states do not directly apply. 
Let us first briefly recapitulate the phase diagram of this system  in the absence of rotation. 
Its main feature is that for significantly large drag $|n_d|>n_c$, the easiest topological defects 
to excite thermally are $(1,1)$ vortex loops. Proliferation of these composite defects leads to a 
state with order only in the phase difference, a so-called  super-counter-fluid 
\cite{Kuklov1,Kuklov2,Dahl1}. 
{In order to estimate the phase stiffness
which is left in the system after the (1,1)
vortex loops proliferate, one has to extract from the Eq. (1) 
the term which depends only on the gradients of the phase difference, which stiffness
is thus unaffected by the proliferation of (1,1) loops.}
The corresponding separation of variables in the presence of rotation is { given by}
\begin{align}
\label{compmodel}
F
=\frac{1}{2}\bigg\{&\frac{\frac{n_1n_2}{m_1m_2}-\frac{n_d}{\sqrt{m_1m_2}}\left(\frac{n_1}{m_2}+\frac{n_2}{m_1}\right)}
      {\frac{n_1}{m_1}+\frac{n_2}{m_2}-\frac{\sqrt{m_1m_2}}{\tilde{\mu}^2}n_d}\times\nonumber \\
&\left(
\nabla\theta_1-\nabla\theta_2-\left(m_1-m_2\right)\vect{\Theta}
\right)^2\nonumber \\
&+
\frac{1}
     {\frac{n_1}{m_1}+\frac{n_2}{m_2}-\frac{\sqrt{m_1m_2}}{\tilde{\mu}^2}n_d}\times \nonumber \\
&\left[
\left(\frac{n_1}{m_1}-\sqrt{\frac{m_2}{m_1}}\frac{n_d}{\tilde{\mu}}\right)\left(\nabla\theta_1-m_1\vect{\Theta}\right)\right.
\nonumber \\
&\left.+
\left(\frac{n_2}{m_2}+\sqrt{\frac{m_1}{m_2}}\frac{n_d}{\tilde{\mu}}\right)\left(\nabla\theta_2-m_2\vect{\Theta}\right)
\right]^2
\bigg\},
\end{align}
where $\tilde{\mu}^{-1}=m_1^{-1}-m_2^{-1}$.
\par
 After proliferation of (1,1) vortices it is  the first term 
in (\ref{compmodel}) which accounts for the only phase stiffness
remaining in the system, and we can renormalize the coefficient of
the second term to zero and discard it.
The complexity of the situation 
arising under rotation is that  along with the (1,1) vortex loops excitations there 
are rotation-induced vortex lines. Vortex loops and lines affect each other's orderings 
and proliferation.  Thus, we may ask (i) what are 
the ordering patterns of rotation-induced vortex lines in the model (\ref{model}),
(ii) how do rotation-induced vortex lines contribute to renormalization of stiffness 
and thus to symmetry breakdown patterns, and (iii) can ordering of the rotation-induced 
vortices signal the presence of a negative drag effect. 
\par
To address these questions, we have performed large-scale MC computations using discretization 
of Eq.~\eqref{model} under rotation, in the Villain approximation \cite{Dahl1}.
Throughout, we use a temperature scale such that the temperature $T$ at which two decoupled
superfluids with equal masses and phase stiffnesses transition to normal fluids is $T=3.3$.
A negative intercomponent drag will tend to increase the temperature at which this 
transition occurs.   
\par
Consider first the simplest limit when $m_1=m_2=1$ and  $n_1=n_2=n$.  
Eq. \eqref{compmodel}  then simplifies to
\begin{align}
\label{model_separated}
F=\left(\frac{n}{4}-\frac{n_d}{2}\right)\left(\nabla(\theta_1-\theta_2)\right)^2
+\frac{n}{4}\left(\nabla(\theta_1+\theta_2)-2\vect{\Theta}\right)^2.
\end{align}
When $n_d=0$, the condensates are decoupled and a rotating system forms two hexagonal lattices of 
the types (1,0) and (0,1). For $n_d < 0$, an attractive interaction between rotation-induced 
vortices results. Thus, in the ground state the system  forms a triangular lattice of (1,1) 
vortices.  Such a configuration minimizes  the gradients in the first term in 
Eq. \eqref{model_separated}. In the simplest limit $m_1=m_2=1$ and  $n_1=n_2=n$ we have found 
regimes where the vortex lattice melts while vortices nonetheless retain their composite 
character. The system retains a symmetry in the phase difference thereby representing a 
``rotation-induced" super-counter-fluid state. Introducing a mass and density disparity 
$m_1\neq m_2$ and $n_1\neq n_2$ gives an entirely different ordering and symmetry breakdown. 
This is the main focus of this paper.
\par
We study the spatial symmetry breakdown pattern  and the effect of thermal 
fluctuations, by computing real space averages of vortex densities. They are 
produced by integrating the $z$-directed vortex segments along the $z$-axis, 
$\tilde{\nu}_i(\vect{r}_\perp)=L_z^{-1}\sum_z\nu_i^z(\vect{r}_\perp,z)$, with a subsequent 
averaging over typically $10^4$ different configurations at a given
temperature. 
$\nu_i^z(\vect{r}_\perp,z)$ is the vorticity of component $i$ in the $z$-direction at
$\vect{r}=(x,y,z)$ and $\vect{r}_\perp=(x,y)$ is the position in the
$xy$-plane. Thus, for an elementary vortex on the numerical grid directed along z-axis,
the quantity $\nu_i^z(\vect{r}_\perp,z)$ is nonzero and positive in lattice plaquette which 
corresponds to the center of the vortex. It is nonzero and negative for an antivortex, whence 
$\tilde{\nu}_i(\vect{r}_\perp)$ gives the average $xy$-position density of the 
rotation-induced vortices.
\par
\begin{figure}[h!!] 
  \includegraphics[width=0.7\columnwidth]{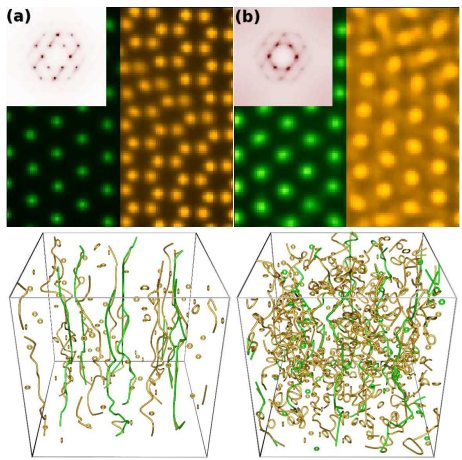}
\caption{(color online) \label{graphene} 
{ Vortex orderings for the parameters  $n_2/n_1=4$, $n_d/n_1=-5.0$ and
$m_2/m_1=2$.} Panel~(a) shows the low temperature phase, { ($T=6.90$)},
where vorticity averages { (shown 
in gold for component $2$ and  green for component $1$)} produce a hexagonal 
vortex lattice in one component and a honeycomb lattice in the other component. 
Panel~(b) shows the situation occurring at a higher temperature { ($T=9.52$)}, 
where a hexagonal sublattice of the honeycomb lattice melts, the number of vorticity 
peaks in real-space averages diminishes by a factor two, and the remaining hexagonal 
lattice is co-centered with a vortex lattice of component $1$. The the insets in the 
top panels in part (a) and (b) show the structure factor evolution. The lower panels
show  a typical $3d$ vortex configuration in a 
$24\times 24\times 24$ segment of the system.}
\label{honeycomb}
\end{figure}

 
Let us consider the 
case $n_2/n_1=4$, $n_d/n_1=-5.0$ and $m_2/m_1=2$. 
{ Now, since the vortex density is proportional to $\Omega m_i$ \cite{Feynman_Stat_Mech_Book},  
there are twice as many vortices in component $2$ as in component $1$ for these parameters.} 
The system  exhibits a striking vortex ordering. Component $1$ forms a triangular lattice, while 
component $2$ with twice as many vortices, forms a honeycomb lattice. Every second vortex in 
the honeycomb lattice is co-centered with a vortex of the other component. This can be also viewed as
an ordered equal mixture of (1,1) and (0,1) vortices. We find that the structure with a 
honeycomb plus hexagonal vortex lattice persists for a significant range of temperatures. 
Fig. \ref{honeycomb}a shows a real-space average and a $3d$ snapshot of a typical configuration 
of this spatial symmetry breakdown pattern at $T=6.9$.  This ordering has broken down at $T=9.5$, 
where we observe a partial meltdown manifested in the disappearance of every second vortex 
position peak in the real-space averages. However, every other vorticity peak corresponding 
to a hexagonal sublattice co-centered with vortex lattice of component $1$ survives. 
The reduction of the number of vorticity peaks in component $2$, the corresponding change in 
the structure factor, along with a $3d$ snapshot of a typical vortex configuration, is shown 
on Fig. \ref{honeycomb}b. The structure factor of component $i$ is defined as
$S^{(i)}(\vect{k}_\perp)=\left|{{1}/{(L_xL_yf)}\sum_{\vect{r}_\perp}\tilde{\nu}_i(\vect{r}_\perp)\ex{-\im\vect{r}_\perp\cdot\vect{k}_\perp}}\right|^2.$

\begin{figure}[h!!] 
  \includegraphics[width=0.7\columnwidth]{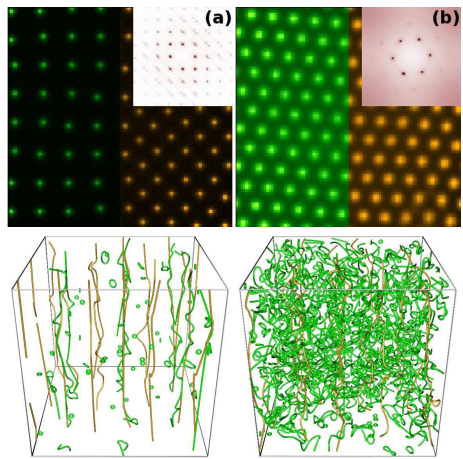}
\caption{(color online)  { Vortex orderings for the parameters
 $n_2/n_1=16$, $n_d/n_1=-2.5$ and
$m_2/m_1=2$. }
The $\vect{k}$-space inset is the full $\vect{k}$-space structure factor for component $1$, 
in both panel~(a) and (b). The left hand side in panel (a) and (b) is a realspace average 
{ of component $2$ (yellow), while the right hand side is component $1$ (green). }
Column (a) corresponds to the 
ordering in low-temperature {($T=11.0$).
Coulumn (b) shows the situation taking place at higher temperature {($T=13.3$)}
where  the green vortices (component $1$) are in the phase corresponding to a dual superfluid 
bosonic density wave.}
 It is far from obvious from the $3d$ picture in the panel (b) that component 
$1$ breaks translational symmetry. However, when averaged  over many configurations,  
we see that component $1$ indeed breaks translational symmetry.}
\label{SS}
\end{figure}

Let us finally turn to the case where $m_2/m_1=2$, but $n_2/n_1=16$. Now, with  
$n_d/n_1 =-2.5$, we find that at low temperatures, the system instead forms 
two square lattices. 
In the ground state the square lattice in component $2$, which has twice as many vortices 
as component $1$, is rotated $45$ degrees with respect to the lattice of component 
$1$, so that every second vortex is co-centered with a vortex of component $1$ 
lattice, see Fig. \ref{SS}a. { Again this can be viewed as an equal mixture of (1,1)
and (0,1) vortices.}  Note that, 
 in contrast to 
the case of two-component vortex matter with only repulsive interactions \cite{Mueller-Ho,Dahl2},
here  the vortex lattices are not interlaced. Here, the appearance of square symmetry 
is caused by attractive interactions between vortices of different types.

\par
{ As the temperature is increased from $T=11.0$ to  $T=13.3$, cf. Fig. \ref{SS}, 
the evolution of the system  is particularly remarkable: we observe a discontinuous phase transition, where in 
the real-space averages the number of vortex position peaks in component $1$ {\it doubles}. }
This should be compared with the previous case of lower { disparity of stiffnesses. There, in 
contrast, the system undergoes a transition to a state where vortex position peaks of the 
other component  was reduced by a factor of two}. Furthermore, both lattices change symmetry 
by collapsing onto a hexagonal co-centered configuration, as seen in the right panel of 
Fig. \ref{SS}. A $3d$ snapshot of a part of the system, shown in the lower panel in Fig. 
\ref{SS}, reveals that the process is accompanied by a rapid increase of vortex loops in 
component $1$. Furthermore, the figure \ref{SS2} shows the central feature of this state. 
Namely, the helicity modulus, {  equivalently the superfluid density computed according 
to the procedures in Ref. \cite{Dahl1}}, for component $1$ disappears essentially simultaneously 
with the structure factor for the square lattice in component $1$.  However, at the same time  
there emerges a nonzero triangular structure factor in component $1$. It extends for a significant 
range of temperature {\it where the helicity modulus of component $1$ is zero}. 
Therefore, the above observations are { not} related to a standard vortex-loop proliferation 
transition in the $3dXY$ universality class. If this behavior were associated with a 
standard vortex-loop proliferation transition of the vortices in component $1$, the superfluid 
stiffness (helicity modulus) of this component would vanish simultaneously with the structure 
factor of the corresponding vortex lattice. 
\par
Thus, we have a quite remarkable situation. 
On the one hand, zero helicity modulus in z-direction indicates that vortices 
are entangled with each other like in a vortex liquid, a state
which has a dual counterpart in superfluid bosons \cite{MPAFisher}. On the other hand, the 
vortex system nonetheless features a structure function which is characteristic of a vortex 
lattice, namely it has distinct peaks at reciprocal lattice vectors.  Thus, the dual counterpart of the vortex system 
we found, is a bosonic superfluid density wave. 
\par
In terms of vortex matter this corresponds to the following situation. Vortices in component $1$ 
in this state are largely co-centered with the vortex lattice of component $2$,  but at these 
temperatures constantly and  freely switch from being co-centered with one to being co-centered 
with the another vortex at different points along the $z$-axis. {In order for the number of vortex 
position peaks of component 1 to be double the number that is generated by the rotation, a large 
number of (1,0) vortex loops must be induced. Then the part of each (1,0) loop that is parallel 
to the (0,1) vortices have a tendency to be co-centered with a (0,1) vortex, breaking translational
symmetry for this segment, while the remaining part of the (1,0) loop which is not parallel  to 
the (0,1) vortices has a random position and does not break translational symmetry.
Only at further elevated temperatures, a crossover takes place where vortex loops proliferate, 
and the vortices  loose line tension and the structure 
factor vanishes, see Fig. \ref{SS2}.
\par

\begin{figure}[h!!]
  \includegraphics[width=0.5\columnwidth]{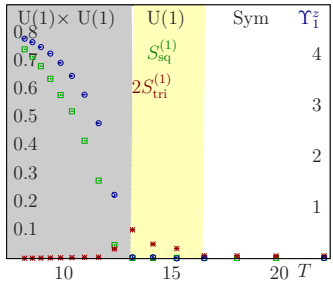}
\caption{(color online) 
Plot of the evolution of the structure factor for component 1 (red and green)
when with increasing temperature the system undergoes a sequence of the phase transitions from the state
with $U(1)\times U(1)$ broken symmetry  to the state with $U(1)$ broken symmetry
and finally to the fully symmetric state (denoted as ``Sym").
 The (green) squares represent 
the structure function of the {\it square lattice} in component 1, $S_{sq}^{(1)}$, at  
the Bragg-peaks at  $\vect{k}_\perp=(0.7854,0.00)\approx(\pi/4,0)$. The (red) crosses 
represent the structure function of the {\it triangular lattice} in component 1, $S_{tri}^{(1)}$,
at the Bragg-peak at $\vect{k}_\perp=(1.1781,0.0982)\approx(3\pi/8,\pi/64)$.  Furthermore, 
the (blue) circles represent the helicity modulus in the $z$-direction for component 1. It 
vanishes at the same temperature as the square lattice ordering ceases. However
at the same temperature there appears a nonzero structure factor for a  triangular 
lattice $S_{tri}^{(1)}$. This is accociated with the vortex state dual 
to bosonic superfluid density wave. The parameters
are $n_2/n_1=16$, $n_d/n_1=-2.5$ and $m_2/m_1=2$. The system size is $L\times L \times L$, 
with $L=64$. Periodic boundary conditions are used in all directions, $10^5$ sweeps are 
used for thermalization and $10^6$ sweeps for collecting average values with sampling 
every $100^{\rm th}$ sweep.}
\label{SS2}
\end{figure}
In conclusion,  we have considered two-component superfluids with a negative dissipationless 
drag.
In the model  Eq. \eqref{model}, the  underlying optical lattice plays only a microscopic role 
by providing a negative intercomponent drag through the mechanisms discussed in \cite{Kuklov1}. 
{ Thus , Eq. (\ref{model}) describes the system at  temperatures T larger than the 
vortex pinning energy of optical lattice $E_p$ and thus there 
is no lattice pinning effect \cite{Cornell}. The vortex ordering pattern in this system 
is strongly affected by the negative dissipationless drag, resulting in the formation of  square 
and honeycomb lattices. Observation of these different ordering symmetries in experiments
would be the hallmark of intercomponent drag.} At finite temperature there are phase transitions between 
states with different lattice symmetries. The main conclusion of our paper is that, apart from 
different patterns of spacial symmetry breakdown, the standard notions of vortex ordering single-component 
vortex matter do not directly apply in the case of two-component vortex matter with a negative drag. 
Namely, we have identified a state of 
vortex matter which is dual to a bosonic superfluid density wave, where one of the components breaks 
translational symmetry even though there is {\it no} symmetry broken in the order parameter space.
In this regime,  a standard experimental technique of a density snapshot with a significant averaging 
along the $z$-axis would indicate a vortex lattice  even though this is not a superfluid state. Since 
this state is phase-disordered, it can be discriminated from superfluid vortex lattice via interference 
experiments. In an experimental situation, these effects will be naturally affected by density inhomogeneities 
present in traps. However, studies \cite{trap} of the effect of the presence of a trap on three dimensional vortex matter,
 suggest that the above states can be realizable in an extended area near the center of the trap.
\par
This work was supported by the Norwegian Research Council Grants No. 158518/431 and No. 158547/431 (NANOMAT), 
and Grant No. 167498/V30 (STORFORSK).

\end{document}